\documentclass{emulateapj}

\newcommand{\kms}{km~s$^{-1}$}
\newcommand{\msun}{{\it M}$_{\odot}$}

\newcommand{\etal}{{\it et al.}}

\newcommand{\be}{\begin{equation}}
\newcommand{\ee}{\end{equation}}
\newcommand{\bi}{\begin{itemize}}
\newcommand{\ei}{\end{itemize}}

\shorttitle{Impact of Distance Uncertainties on LFs}
\shortauthors{Masters \etal}

\begin{document}

\title{The Impact of Distance Uncertainties on Local Luminosity and Mass Functions}
\author{Karen L. Masters\altaffilmark{1}, Martha P. Haynes\altaffilmark{1,2} and Riccardo Giovanelli\altaffilmark{1,2}}
\altaffiltext{1}{Center for Radiophysics and Space Research, Cornell University, Space Sciences Building, Ithaca, NY 14853; masters@astro.cornell.edu, riccardo@astro.cornell.edu, haynes@astro.cornell.edu}
\altaffiltext{2}{National Astronomy and Ionosphere Center, Cornell University, 
Space Sciences Building, Ithaca, NY 14853. 
The National Astronomy and Ionosphere Center is operated by Cornell University 
under a cooperative agreement with the National Science Foundation.}

\begin{abstract}
In order to investigate discrepancies between recent published estimates of the
the HI mass function (HIMF), we explore the impact of distance uncertainties on the derivation
of the faint end slope of mass and luminosity functions of galaxies in the local volume by deriving HIMFs from
mock HI surveys. We consider various survey geometries and depths and compare the HIMFs measured when using 
``real'' distances, distances derived by assuming pure Hubble flow and distances assigned from 
parametric models of the local velocity field.  The effect is variable and dependent on the exact survey geometry, but can easily lead to incorrect 
estimates of the HIMF, particularly at the low mass end. We show that at least part of the discrepancies 
among recent derivations of the HIMF can be accounted for by the use of different methods to assign
distances. We conclude that a better understanding of the local velocity field will be necessary 
for accurate determinations of the local galaxy luminosity and mass functions.
\end{abstract}


\section{Introduction}

One of the perplexing discrepancies between observations and CDM theory revolves
around the ``missing satellite'' problem.
Numerical simulations in $\lambda$CDM predict a value for the logarithmic slope of the faint end of the halo mass function, $\alpha$, which is 
close to the value of $-1.8$ 
that arises analytically from the Press-Schechter formalism (Press \&
Schechter 1974), but most recent determinations of the optical 
luminosity function (LF) in the local volume 
yield values of $\alpha$ that are significantly flatter.
Complementary to the results on the optical LF, several determinations of the
local HI mass function (HIMF) have likewise yielded relatively flat faint-end slopes.
The over-prediction of the number of low mass objects relative to those actually observed 
is considered one of the last remaining stumbling blocks for $\lambda$CDM, and points 
towards a new understanding of the baryon physics of galaxy formation. It is obviously 
desirable that all of the uncertainties in the observed luminosity and mass functions 
be well understood.

To derive a luminosity (or mass) function from observational data, the sample selection effects 
must be considered carefully. In most surveys, the lowest luminosity 
(or mass) objects are only visible nearby, and their contribution must be 
weighted accordingly. 
Fractional distance errors can be large for relatively nearby
objects, and can thus have a very strong impact on determinations
of the faint end characteristics of luminosity (or mass) functions.
When those functions are evaluated using data obtained in clusters,
the principal source of uncertainty lies in the assessment of
membership; assuming that uncertainty is to first order independent
on luminosity (or mass), it can have a small impact on the estimate
of LF slopes. Determinations of the
HIMF, however, have to rely on relatively shallow, wide area surveys to detect low HI mass objects over a sufficient volume, and are thus most susceptible 
to uncertainties associated with distance errors in the nearby population.

Primary distances (eg. cepheids, the tip of the red giant branch or surface brightness fluctuations, the best of which give distances to $\pm$ 10\%) are available for
$\sim 600$ galaxies, most within $\sim 30$ Mpc. Secondary distances (which rely on the above methods for calibration) carry larger
errors ($\sim$ 15--30\%) but do account for peculiar velocities for some 5000 objects.
For all other galaxies, the only indication of distance is their redshift. 
Outside of $cz=6000$ \kms, pure Hubble flow works well, since a typical peculiar velocity of a few 
hundred \kms ~creates an uncertainty on the redshift distance of $<$ 10\%. For closer 
galaxies, however, the relative error is much larger, and, for the nearest objects, whose recessional velocities are dominated by peculiar motions, the error can be a factor of 
two or more. Furthermore, peculiar velocities are not random but 
exhibit coherent flows 
towards local over-densities. In some directions, the redshift-distances of all (nearby) galaxies  
will be systematically under- or over- estimated if peculiar 
velocities are ignored. 

The effects of peculiar velocities on distance determinations
can be mitigated by the use of parametric models of the local velocity field. While 
clearly oversimplifications of the real flows, such models give a direction-dependent 
redshift-distance prescription. A recent example of such a model is the multi-attractor 
model of Tonry \etal ~(2000; hereafter TB00: also see Willick \& Batra ~2001).

In order to investigate the effects of distance uncertainties   
on local luminosity or mass functions, we consider here the derivation of the 
HIMF. Recent determinations of the local HIMF have yielded discrepant results on 
the abundance of low HI mass objects, namely: Zwaan \etal ~(1997; Z97), Rosenberg \& 
Schneider (2002; RS02) and Zwaan \etal ~(2003; Z03). The Z03 HIMF is based on the 
HI Parkes All Sky Survey (HIPASS) Bright Galaxy Catalogue (BGC), while the Z97 and RS02 HIMFs are both based 
on drift scan surveys conducted at Arecibo during the period of its recent
upgrade (respectively called AHISS, the Arecibo HI Strip Survey, and ADBS,
the Arecibo Dual Beam Survey).  
A summary of the results of those surveys
is provided in Table 1, parameterized as Schechter functions, 
\be \phi(x) = \frac{dn}{d\log M} = \ln 10 ~\phi_\star~ x^{(\alpha+1)}~ e^{-x},
\ee 
where $x=M/M_\star$, $M$ is the HI mass, and $n$ is the total number density of galaxies. 
The faint end 
slopes of the HIMF derived by the three surveys vary 
between $-1.20$ and $-1.53$, yielding extrapolations below $M_{HI} = 10^7$
\msun~ that disagree by an order of magnitude, the RS02 HIMF having the
steeper slope. 

\begin{deluxetable}{lccc}
\tablecaption{Parameters in published HIMFs}
\tablewidth{0pt}
\tablehead{
\colhead{Reference} &\colhead{Z97} & \colhead{RS02} & \colhead{Z03} \\
\colhead{Survey} &\colhead{AHISS} & \colhead{ADBS} & \colhead{HIPASS} }
\startdata
$\phi_\star [h_{70}^3 {\rm Mpc}^{-3}]$ & 0.0048 & 0.0047 & 0.0070  \\
$\log(M_\star/M_\odot) + 2 \log h_{70}$ & 9.86 & 9.94   & 9.85  \\
$\alpha$ & -1.20 & -1.53  & -1.30  \\
Survey area & 65 deg$^2$ & 430 deg$^2$  & 20,600 deg$^2$ \\
$D_{\rm lim}$ for $10^8$\msun & 100 Mpc & 13 Mpc & 7 Mpc \\
\# galaxies & 61 & 265 & 1000 \\
\enddata
\end{deluxetable}

Previous studies of the impact of distance errors (from ignoring peculiar velocities) on HIMFs have modeled the effect, often called the `Eddington effect', 
by adding random Gaussian noise to the distances. Using this model, RS02 conclude that 
the effect is negligible in their survey volume until velocity dispersions of 600 \kms ~are 
reached, while Z03 find that for an assumed velocity dispersion of 50-100 \kms ~in the 
local universe ($cz < 3000$ \kms) the slope of the low mass end of the HIMF probably 
steepens by $\Delta \alpha < 0.05$. Since the peculiar velocity field in the local universe 
is not random but rather dominated  by in-fall onto the Virgo cluster, these models 
could easily hide the main bias. In this paper, we investigate whether all or part of the 
discrepancy among the three HIMF determinations can be accounted for by the effects of 
distance uncertainties introduced by the different methods used by Z97, RS02 and Z03 to 
assign distances. 

Here, we construct mock catalogs designed to mimic the HI surveys conducted by Z97, RS02 
and Z03 and then derive mock HIMFs under different distance assumptions. 
$H_\circ = 70 ~{\rm km~s}^{-1}{\rm Mpc}^{-1}$ is assumed throughout, and 
the dependence on $H_\circ$ is explicitly listed using $h_{70} = 
H_\circ/(70~ {\rm km~s}^{-1}{\rm Mpc}^{-1}$). 

\section{Survey Simulations}

We first construct a mock catalog of HI rich objects in the local universe, 
which is then ``observed''.
 In order to account for the known local variations in 
the 3D distribution of galaxies, a version of the density and velocity fields derived 
from the IRAS Point Source Catalog Redshift Survey (PSC$z$) with a value 
$\beta = \Omega^{0.6}_{mass}/b = 0.5$ is used (kindly provided by E. Branchini, see 
Branchini \etal ~1999 for details). This map extends to a distance of 120 $h^{-1}$ Mpc 
with a grid spacing is 0.9375 $h^{-1}$ Mpc in the inner 60 $h^{-1}$ Mpc, and twice that 
in the outer shell. The map has been smoothed with a Gaussian filter of radius 3.2 $h^{-1}$ Mpc.

The volume is seeded with HI rich galaxies according to the density field, and masses 
are assigned 
 from a given HIMF. The HI sizes of the galaxies 
and velocity widths 
are calculated according to empirical scaling 
relations derived from our own HI survey data and from Broeils (1992).
Inclinations are randomized, realistic scatter is added to the scaling relations (as measured from the empirical fit) and a turbulent velocity of 10 \kms ~is added 
to construct the observed velocity width.

The HI mass of a galaxy at a distance $D$ (in Mpc), and with observed HI line flux 
can be approximated as 
$M_{HI}/M_\odot \simeq 2.356 \times 10^5 D^2 S_{\rm peak} W_{\rm HI},$
where $S_{\rm peak}$ is the peak flux in the HI line in Jy and $W_{\rm HI}$ is its width in \kms.
The detectability of a HI rich galaxy depends on both the peak flux and the width of the 
HI line. If we assume that smoothing of the spectrum is applied to match the width of 
the lines, up to a maximum of $W_{\rm smo}=100$ \kms, the signal-to-noise ratio can be derived
from the standard radiometer equation such that
\be
S/N  =  \frac{6.06 \left( \frac{M_{HI}}{10^6 M_\odot} \right ) 
\left ( \frac{W_{\rm HI}}{100} \right )^\gamma f_\beta t_{\rm int}^{1/2} \sqrt({\rm CBW}/25 {\rm kHz})}{(D^2 T_{\rm sys}/G)},
\ee
where $\gamma =-3/4$ for $W_{\rm HI} \leq 100$ \kms, and $\gamma= -1$ for 
$W_{\rm HI} > 100$ \kms, accounting for the effects of smoothing; $f_\beta$ 
is the fraction of the source covered by the telescope beam; $t_{\rm int}$ is 
the integration time; CBW is the channel bandwidth and $T_{\rm sys}/G$ describes 
the gain of the telescope. Specific 
details of the simulated observations are given in Section 3 below.

The observed HI mass of a galaxy is
$M_{\rm obs} \nolinebreak  =  \nolinebreak M_{\rm true} (D_{\rm obs}/D_{\rm true})^2,$
where $D_{\rm obs}$ is the distance assigned to the galaxy and $D_{\rm true}$ 
is its true distance. From the set of ``detections'' we construct HIMFs by 
summing $1/V_{\rm lim}$ (the total volume within which a galaxy can be seen) weighted 
by the average density in that volume (from the PSC$z$ density field) in bins 
of $\log (M_{HI}/M_\odot)$. This method prevents over- or under- dense regions 
from giving an over- or under-estimate of number counts in certain bins.
If $D_{\rm obs} = D_{\rm true}$, we recover the input HIMF. We investigate below 
the effect that $D_{\rm obs} \ne D_{\rm true}$ has on the recovered HIMF for 
various survey geometries and depths.

\section{Results and Discussion}

\subsection{Simulated HIPASS BGC}

First, we consider a mock catalog constructed to mimic the HIPASS BGC
which contains the 1000 galaxies in 
the southern sky with the largest peak HI flux (see Z03). To remove the effects 
of cosmic variance an average obtained from 10 simulations is used. 

Most of the galaxies in the real BGC have their distances estimated using Hubble's 
Law, and yet are within $cz=3000$ \kms. We investigate what effect this assumption has (if any) 
on the derived HIMF. As noted in Table 1, the HIMF derived from the HIPASS BGC has a 
shallower low mass end slope than that derived from the ADBS in RS02. We will see 
if any part of this discrepancy can be accounted for 
by the use of Hubble's Law to assign distances by Z03. 

\begin{figure}
\epsscale{1}
\plotone{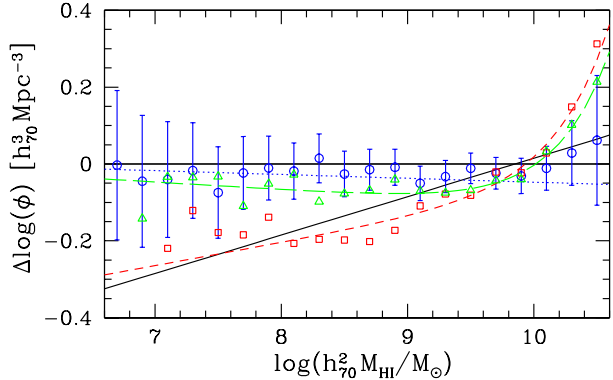}
\caption{\label{HIPASS} Difference between input and reconstructed HIMFs for a mock HIPASS BGC survey (shown are the average 
of 10 simulations).  The HIMF derived when ``true'' distances are used (circular points and dotted line), along with the HIMF constructed when multi-attractor model distances are used (triangular points and long-dashed line) reconstructs the input HIMF within the errors (here typical Poisson counting errors are shown only on the ``true'' distance points). When pure Hubble flow is assumed (square points and short dashed line), the low mass end of the HIMF is underestimated. The solid line shows the Z03 HIMF, which differs from the input by $\Delta \alpha = 0.1$. }
\end{figure}

Figure \ref{HIPASS} shows 
the difference between the input and reconstructed HIMFs for the input (or ``true'') 
distances, distances derived by assuming Hubble flow, and distances from the 
multi-attractor model of TB00. Clearly, significant variations arise among the HIMFs 
constructed from the mock HIPASS BGC when different distance estimators are used. 
The HIPASS survey volume does not contain the Virgo Cluster 
and therefore has a fairly quiet local velocity field; however, as can be seen 
from the figure, the low mass slope of the HIMF will be {\bf underestimated} if 
pure Hubble flow is assumed. In Figure \ref{HIPASS}, the input HIMF is the same as Z03 (see Table 1), except that the low mass slope has been steepened to 
$\alpha = -1.4$. The HIMF recovered from this mock HIPASS BGC is close to that reported 
by Z03 when Hubble flow is assumed.

We suggest that an adjustment of $\Delta \alpha \simeq -0.1$ should be applied 
to the Z03 low mass slope, making it $\alpha = -1.4 \pm 0.1$. This correction makes the result {\bf consistent} with the RS02 result of $\alpha = -1.53 \pm 0.2$ (where here the error on the RS02 result is estimated by scaling that reported for Z03 by $\sqrt N_{\rm gal}$). This change in low mass slope, while interesting in the light of comparisons with $\lambda$CDM, has little effect in the total HI mass density implied by the surveys. The total mass density, dominated by $M_\star$ galaxies, is already in agreement between the two surveys.

\subsection{Simulated Virgo cluster survey}

As the nearest rich cluster to us, Virgo represents the highest density region in the 
local Universe, and thus constitutes the obvious choice for the study of the HIMF in a 
high density environment. The northern hemisphere sky accessible to the Arecibo telescope 
covers both the Virgo cluster and comparably-far regions of lowest cosmic density 
located roughly in the anti--Virgo direction. The arrival in 2004 of the Arecibo L-Band Feed Array 
(ALFA) will revolutionize mapping capability at the Arecibo telescope and will allow
study of possible variations in the HIMF in different density environments, already suggested
by RS02 but with low statistical significance. Here, we investigate the potential
of a proposed ALFA survey of an area $30^\circ \times 30^\circ$ centered on the Virgo 
cluster (RA=$11-13^h$, DEC=$0-27^\circ$), assuming $T_{\rm sys}/G= 3$ Jy, and 
with $t_{\rm int} = 12$s (drift scan) and CBW=25 kHz. The mean noise in such a survey would be $\sim$ 4 mJy/beam (at our assumed maximum resolution of 5 \kms). This can be compared to 20 mJy/beam for HIPASS and $\sim$ 8.6 mJy/beam for the ADBS, see \citet {BS01} and RS02; here both values have been scaled to a resolution of 5 \kms.
We also consider an equal area in the anti-Virgo region. (RA=$23-1^h$, DEC=$-27-0^\circ$). This 
region of the sky is not accessible to Arecibo, but for the purposes of illustrating 
the effects of distance uncertainties on the construction of HIMFs, the combination of 
such a Virgo-anti-Virgo survey provides a good comparison between the nearest regions of high 
and low cosmic density.

\begin{figure}
\epsscale{1}
\plotone{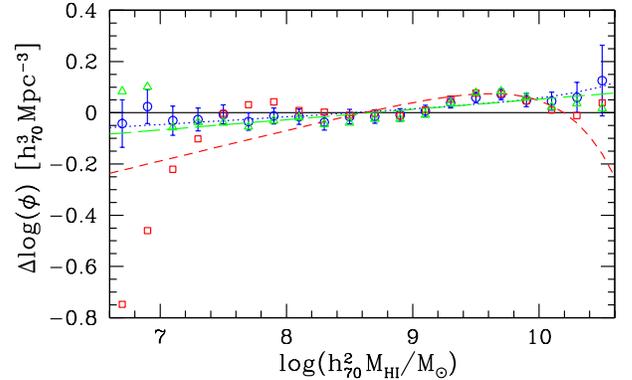}
\caption{\label{virgo} Difference between input and reconstructed HIMFs from simulated ALFA drift scans of Virgo 
(shown is the average of 10 simulations; roughly 4000 galaxies are seen in each survey). 
As in Figure \ref{HIPASS}, the input HIMF (that of Z03, but with $\alpha$ steepened to -1.4) has been subtracted, but note that here the 
y-scale has been extended down to accommodate the huge deviation when Hubble flow distances 
are used. Line and point types are the same as in Figure \ref{HIPASS}. 
When the ``true'' distances are used the fit to the low mass slope 
of the HIMF is $\alpha = -1.37$; the multi-attractor model distances give $\alpha = -1.36$, 
while the Hubble flow distances give $\alpha = -1.28$.}
 \end{figure}

\begin{figure}
\epsscale{1}
\plotone{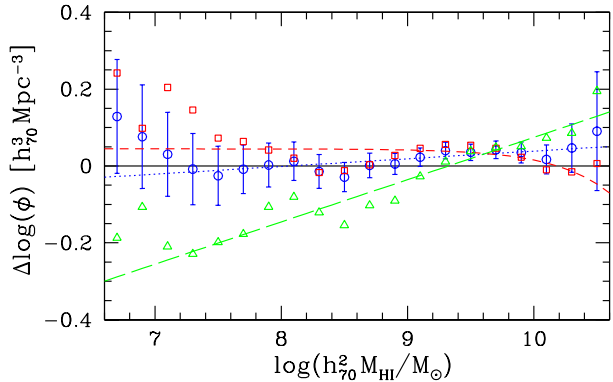}
\caption{\label{avirgo} Difference between input and reconstructed HIMFs from simulated ALFA drift scans of the Anti-Virgo region (shown is the average of 10 simulations; roughly 3000 galaxies are seen in each survey).  Line and point types are the same as in Figures \ref{HIPASS} 
\& \ref{virgo}. Note that this region will not be accessible to ALFA, but is included 
here to illustrate the difference between comparable surveys of the nearest regions 
of high and low cosmic density.}
\end{figure}  

In the Virgo region, we are able to reconstruct the input HIMF well when the ``true'' 
distances are used, and reasonably well with distances taken from the multi-attractor 
model of TB00. However, if pure Hubble flow is assumed, the number of galaxies at the 
low mass end of the HIMF is seriously underestimated (see Figure \ref{virgo}). 
The lowest mass galaxies in our mock survey are seen only in the foreground of Virgo 
(the mass limit at 17 Mpc, the approximate distance of the Virgo cluster, is $\sim 10^8$ M$_\odot$ from  Eqn 2 assuming W$_{\rm HI}$=100 \kms ~and CBW=25 kHz).
If pure Hubble flow is assumed the vast majority of these foreground galaxies will have 
their distances (and hence masses for a given HI flux) overestimated, since they are 
in-falling towards Virgo. 

The two recent Arecibo--based derivations of the HIMF (Z97 and RS02) include part of the 
Virgo cluster within the overall survey area. The low mass end slopes derived from these 
surveys differ by $\Delta \alpha = 0.3$ (with Z97 having the shallower slope). In RS02,
the multi-attractor model of TB00 is used to assign distances, while in Z97 pure Hubble 
flow is assumed. However, the difference between the two HIMFs cannot be explained by 
their use of different distance estimators. The Z97 survey was designed to be extremely 
sensitive but covers a very small area. 
In deeper surveys, we expect the effect described above diminishes. We have
constructed a Z97-like survey (of similar depth and area), and, as expected, the HIMF can be 
reconstructed equally well in such a deep survey with all three distance estimates. 
We therefore must resort to the usual arguments of small-number statistics (noting that 
there were only 65 galaxies in the Z97 HIMF) and differing treatments of completeness 
to explain the discrepancy between the HIMFS of Z97 and RS02.

In the Anti-Virgo region, the input HIMF can be reconstructed well using either 
``true'' distances or by assuming Hubble flow. Here the use of the multi-attractor 
model of TB00 causes the largest bias; acting to create an underestimate of the number 
of low mass HI galaxies (see Figure \ref{avirgo}). 
The multi-attractor model of TB00 is based on distances to $\sim$ 300 elliptical galaxies. 
There are approximately 30 of these galaxies in the Virgo region discussed here, but fewer 
than 5 in the anti-Virgo region. At the depth of these HI surveys, the anti-Virgo region is 
relatively under-dense, making it unavoidable that there will be few tracer galaxies in that 
direction. This undersampling has the effect of forcing the model for the velocity field to fit better 
towards Virgo at the expense of the fit in the anti-Virgo direction. The multi-attractor 
model of TB00 consistently over-estimates the distance for a given velocity in the 
anti-Virgo direction leading to the observed underestimate of the number of low mass 
galaxies (as discussed above). 

\section{Conclusions}

We have used mock HI survey data to demonstrate that 
distance uncertainties can create serious biases in the low mass end of the HIMF;
similar arguments could be applied to studies of the field galaxy optical LF.
When the effects of the local peculiar velocity field are ignored, incorrect estimates 
of the HIMF will be derived. The exact nature and size of the bias will depend both on 
the survey geometry and its depth. 

In a survey constructed to simulate the HIPASS BGC, the low mass HIMF slope 
is {\bf underestimated} if peculiar velocities are neglected. We suggest that 
the HIMF reported by Z03 suffers from this problem, which when corrected, would bring 
the Z03 and RS02 HIMFs into agreement within their errors. The adjusted value of the 
Z03 low mass slope is $\alpha=-1.4\pm0.1$, which is to be compared with the RS02 value 
of $\alpha=-1.5\pm0.2$. This change does not affect the implied HI mass density since
it is dominated by $M_\star$ galaxies.

For a survey where low mass galaxies are seen only in the foreground of Virgo,
the distances (and therefore masses) of the nearby low mass galaxies will be overestimated
and the low mass end of the HIMF {\bf seriously underestimated} if in-fall onto Virgo 
is neglected. In deeper surveys, the effect becomes negligible, so that differences in
the adopted distances alone cannot explain the discrepancy between the Z97 and RS02 HIMFs.

A simulated survey in the anti-Virgo region points to the limitations of currently 
available models for the local velocity field. In such a survey, use of distances taken from a multi-attractor model for the local velocity field leads to an underestimate of the low mass HIMF slope. Currently available flow models are not well constrained in the anti-Virgo direction because of a lack of galaxies with primary distances in the region. Velocity field models based on larger 
numbers of tracer galaxies will be required to solve this problem.

The use of a good model for the local peculiar velocity field is found to be essential 
to provide distances for nearby galaxies for which no redshift-independent distances are 
available. Without such a model, it is difficult to derive an unbiased HIMF, and in 
particular, the low mass end may be in error. Disentangling the effects of environment 
and evolution of the HIMF will not be possible unless the local flow model is well 
understood. Other applications where distances to galaxies are needed will also benefit 
from a better understanding of local flows.

\acknowledgements

This research was supported by NSF grants AST-9900695, AST-0307396 and AST-0307661.

\end{document}